# THE AHARONOV–BOHM EFFECT AS A MATERIAL PHENOMENON

V. Rubaev[1] and L. Fedichkin[2]

[1]NIX, Zvezdny blvd., 19, Moscow 129085, Russia

[2]Valiev Institute of Physics and Technology, Russian Academy of Sciences, 34, Nakhimovsky pr., Moscow 117218, Russia; email: leonidf@gmail.com

## ABSTRACT

An experiment to observe the Aharonov-Bohm effect is discussed. A solenoid which consists of a large number of point magnetic dipoles is considered as the source of a vector potential, which acts on a charged particle, and such potential has an electromagnetic field of zero strength in the region of a nonzero vector potential. A detailed microscopic analysis of the change in the quantum state phase of the entire system, namely, a particle and a set of dipoles, reveals the origin of the apparent nonlocality of the action of the vector potential, and shows the locality of the phase change mechanism. An analysis of an experiment with a solenoid shielded by a superconducting shell is given.

## INTRODUCTION

The current level of techniques for creating and manipulating quantum structures [1, 2] has provided a new impetus to research in the field of creation and practical study of quantum systems and phenomena, which were formerly proposed by the pioneers of quantum physics. Among such proposals is the experiment proposed by Aharonov and Bohm [3, 4] to change the phase of the quantum state of a charged particle which passes a solenoid in the region where the solenoid's magnetic field is negligible. The idea that a quantum particle moving in regions where there are no fields of any kind, along different paths, could nevertheless acquire a different phase, depending on their motion trajectory, seemed at that time surprising.

In this paper we propose an alternative method of phase calculation, which leads to the result of Aharonov and Bohm [3–5], but also eliminates the intrigue of nonlocality from the physical consideration of this phenomenon. The Aharonov-Bohm effect has been intensively studied, and in a number of papers attempts have been made to derive it within the framework of the local theory, for example [6–8]. Vaidman [6], in particular, devoted his work to searching for the local physical sources of this effect. He put forward certain qualitative considerations as to why the seemingly nonlocal action of a vector potential should arise from local interactions, if the entire system of a solenoid and a particle be considered consistently from a quantum-mechanical standpoint.

In [7] an approach was developed to describe the nonlocality of the interaction between a particle and a magnetic field source in the framework of quantum electrodynamics. The particle was allowed to move through a screen with two narrow slits along two predetermined trajectories to the right and left of the solenoid. The authors made calculations under perturbation theory, which were accurate to second-order terms. It was shown that the contribution of local exchange of virtual photons between the particle and the solenoid may well cause the phase difference which arises during the bypassing of the vector potential source from different sides.

In [8], the interaction was also calculated in the presence of a quantum electromagnetic field which physically transfers quantum entanglement from a particle to a solenoid.

A gauge-invariant consideration of the evolution of a particle's state as a function of its motion was proposed. Likewise it was shown that it is possible experimentally to study phase effects without closing the trajectories around the solenoid. However, the choice in [7] and [8] of those variables which describe the electromagnetic field as independent quantum degrees of freedom only increases the number of calculations, which makes a consistent consideration difficult, although in itself it is not necessary for a "local" explanation of the Aharonov-Bohm effect.

In [6], Vaidman offers qualitative considerations, but does not provide a detailed analysis of what is taking place. None of these papers provides a detailed analysis of the problems associated with a specific, material structure of the solenoid. The aim of this paper is to eliminate these deficiencies. We consider the solenoid as a composite object consisting of a large number of point magnetic dipoles. A detailed microscopic analysis of the phase change in the quantum state of the entire system, namely, a particle and a set of dipoles, reveals the origin of the apparent nonlocality in the action of the vector potential, and demonstrates the locality of the phase change mechanism.

In addition, on the basis of our approach we analyse some experiments devoted to testing the Aharonov-Bohm effect [9–13], and show that in some experiments with complete shielding of the vector potential source [10, 11, 13], the Aharonov and Bohm phase may not appear.

## Description of the particle-solenoid system

Consider a point particle with a charge $e$ and mass $m$, which moves in the vicinity of the vector potential source (see Fig. 1) in the region where the magnetic field of the same source is negligible. The Lagrangian of a particle with charge $e$ and mass $m$, which for convenience we will consider as spinless, in a magnetic field with vector potential **A** acquires an addition $\frac{e}{c}(\mathbf{A}\dot{\mathbf{x}})$, where $\dot{\mathbf{x}}$ is the particle velocity. If a magnetic field is created by an infinitely long and infinitely thin solenoid, then the entire set of paths along which a particle can move from the point $x_1$ to the point $x_2$ is divided into two classes – paths which bypass the solenoid "to the left" and "to the right," that is trajectories passing clockwise and counterclockwise, respectively [14]. By performing functional integration over the transition paths to calculate the amplitude $U(x_1, x_2)$ of a particle's transition from point $x_1$ to point $x_2$, we can split the integral into two parts, corresponding to each of these classes. We designate their values obtained in the absence of a magnetic field in the solenoid as $U_R(x_2, x_1)$ and $U_L(x_2, x_1)$.

For a non-zero vector potential, $U_R(x_2, x_1)$ and $U_L(x_2, x_1)$ are substituted by $U_R(x_2, x_1)e^{i\varphi_R}$ and $U_L(x_2, x_1)e^{i\varphi_L}$, where

$$\varphi_{L,R} = \int_{x_1}^{x_2} \frac{e}{c}\mathbf{A}d\mathbf{x}$$

To calculate $\varphi_R$, integration is performed over the path that envelops the solenoid on the right, and for $\varphi_L$ – on the left. Because $\mathbf{H} = \mathrm{rot}\mathbf{A} = 0$ everywhere, save for the area occupied by the solenoid, these values are completely determined by the choice of the direction of bypassing the solenoid, and do not depend on the trajectory of integration.

The scattering amplitude of a particle on the solenoid $f(\theta)$ will be proportional to $\int U(x_2, x_1)\psi(x_1)dx_1$ with an appropriate choice of the initial wave packet $(x_1)$, the upper bound of the time interval $[t_1, t_2]$, on which functional integration is performed, tending to infinity. We then have

$$f(\theta) = f_R(\theta)e^{i\varphi_R} + f_L(\theta)e^{i\varphi_L},$$

and the scattering cross-section will be:

$$\sigma(\theta) = |f_R(\theta)|^2 + |f_L(\theta)|^2 + f_R^*(\theta)f_L(\theta)e^{-i\varphi} + f_R(\theta)f^*_L(\theta)e^{i\varphi},$$

where $\varphi = \varphi_R - \varphi_L = \oint \frac{e}{c} \mathbf{A} d\mathbf{x}$, and integration may be performed over any circuit which encloses the solenoid.

Hence the scattering cross-section will contain a dependence on $\varphi$, which dependence in Aharonov-Bohm's original work is reduced to the factor $\sin^2 \frac{\varphi}{2}$.

The actual view of the cross-section as a function of the scattering angle, which was derived in that work, is not important for us here. However, since 1959 everyone who has studied the Aharonov-Bohm effect has been impressed by the fact that the scattering cross-section is non-zero in the absence of forces which act on the scattered particle. The fact is that the entire magnetic field is contained in an infinitely thin solenoid, which, for the purity of a thought experiment, those who wish to do so can mentally surround with an infinitely high potential barrier. This astonishing fact has been generating confusion in many minds for more than 60 years, and such confusion usually assumes two diametrically opposed forms. Some regard the Aharonov-Bohm effect, which has now been confirmed experimentally, as proof of a certain "reality of the vector potential" [15]. In extreme cases, this comes down to studying the effect of the vector potential in the absence of a magnetic field (sic!) on biological subjects [16–19] and patenting devices for medical treatment using the vector potential. Others, on the other hand, deny the very existence of the Aharonov-Bohm effect (albeit now confirmed experimentally), or alternatively they deny its quantum nature [20, 21], despite the presence of Planck's constant in the original Aharonov-Bohm formula.

Below we suggest an explanation of the Aharonov-Bohm effect, whose numerical results do not depart from the original work, but which does not astonish one with the need for unconditional surrender to the fact that the nature of a particle's motion can change in the absence of any material agents which are capable of influencing such motion.

In other words, we intend to prove that the Aharonov-Bohm effect is caused not only by "the reality of the vector potential," but also by mundane material factors.

The proof scheme is as follows:

1. A solenoid in whose vicinity a genuine Aharonov-Bohm effect can be observed is not a means for demonstrating the reality of the vector potential, but a wholly material body consisting of atoms and electrons, in general – objects to which quantum mechanical consideration can be applied.

2. Further, although the solenoid's magnetic field, as everyone knows, being zero does not act on a passing particle, the magnetic field created by the passing particle does act on the solenoid's components, and surely does.

If the solenoid is considered quantum mechanically, then the scattering amplitude should depend on the variables which describe the solenoid. We further show that this dependence can be reduced to the following:

$$|f(\theta)\rangle = f_R(\theta)|\psi_R\rangle + f_L(\theta)|\psi_L\rangle$$

where $|f(\theta)\rangle$ is used to emphasise the fact that the final state of the "solenoid+ particle" system can now be described not only by its dependence on the particle's scattering angle.

Furthermore, it turns out that the values $f_R(\theta)$ and $f_L(\theta)$ appearing here remain the same as in the "classical" case.

The values $|\psi_R\rangle$ and $|\psi_L\rangle$ here designate the solenoid's states into which it transfers under the influence of the magnetic field created by a classical particle when the particle passes it from the right and left respectively.

We show that in certain special cases $|\psi_R\rangle = e^{i\varphi_R}|\psi_0\rangle$, $|\psi_L\rangle = e^{i\varphi_L}|\psi_0\rangle$, where $|\psi_0\rangle$ is the unperturbed state of the solenoid. In the more general case, there is an asymptotic equality at the number of the solenoid's elementary magnetic dipoles tending to infinity:

$$\langle\psi_L|\psi_R\rangle = e^{i\varphi}$$

By folding the square of the modulus of the scattering amplitude $|f(\theta)\rangle$ of the particle with respect to the solenoid variables so as to calculate the scattering cross-section, we obtain:

$$\sigma(\theta) = |f_R(\theta)|^2 + |f_L(\theta)|^2 + (f_R^*(\theta)f_L(\theta)\langle\psi_R|\psi_L\rangle + c.c.),$$

Taking account of the said asymptotic equality, this coincides with the "classical" scattering cross-section.

The evolution operator of the quantum system $U(q_2, t_2, q_1, t_1)$ transforms the wave function of the initial state $\psi_1(q)$ at time $t_1$ into the wave function of the final state $\psi_2(q)$ at time $t_2$ in the following way:

$$\psi_2(q_2) = (U\psi_1)(q_2) = \int U(q_2, t_2, q_1, t_1)\psi_1(q_1)dq_1 \tag{1}$$

integration is carried out over all variables of the system $q$

The $q$ variables here mean the entire set of variables which describe the system.

The evolution operator is expressed in terms of the Feynman integral:

$$U(q_2, t_2, q_1, t_1) = \int_{(q_1,t_1)}^{(q_2,t_2)} \exp\left(i\int_{t_1}^{t_2} L(q,\dot{q})dt\right) Dq(t) \tag{2}$$

where $L(q, \dot{q})$ is the system Lagrangian.

Consider a free particle of mass $m$, which is described by a vector of spatial co-ordinates $\mathbf{x}$. Its Lagrangian will be

$$L_0(\mathbf{x}, \dot{\mathbf{x}}) = \frac{m\dot{\mathbf{x}}^2}{2} \tag{3}$$

We consider the scattering of a particle with charge $e$ and mass $m$ by an infinite magnetic filament elongated along the $Z$ axis (see Fig. 1).

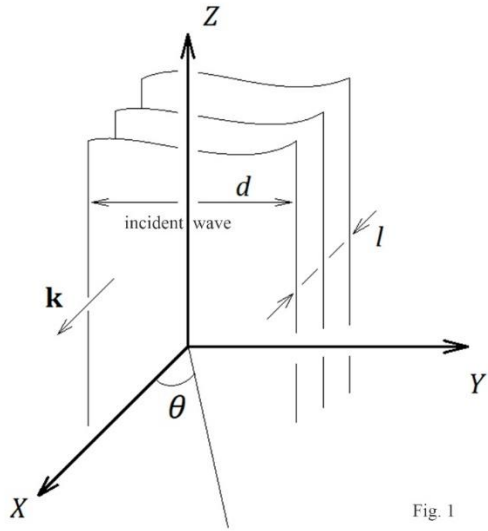

Fig. 1

Let the initial state of the particle be a wave packet of the form $e^{ikx}$, running from minus infinity along the X axis, infinite in both directions along the Z axis, and being of length $l$ along the X axis and of width $d$ along the Y axis (see Fig. 2),

$$\psi_1(x,y) = e^{ikx} u(x) v(y) \tag{4}$$

where $u(x)$ and $v(y)$ are sufficiently smooth envelopes. The final state, for a sufficiently long time $t_2$, will be equal to the sum of that part of the initial wave packet that has passed the filament without experiencing scattering, and an expanding cylindrical wave of the form

$u_1(r) \frac{f(\theta)}{\sqrt{r}} e^{ikr}$, where $r^2 = x^2 + y^2$.

It is evident that in the absence of a filament, $f(\theta) = 0$.

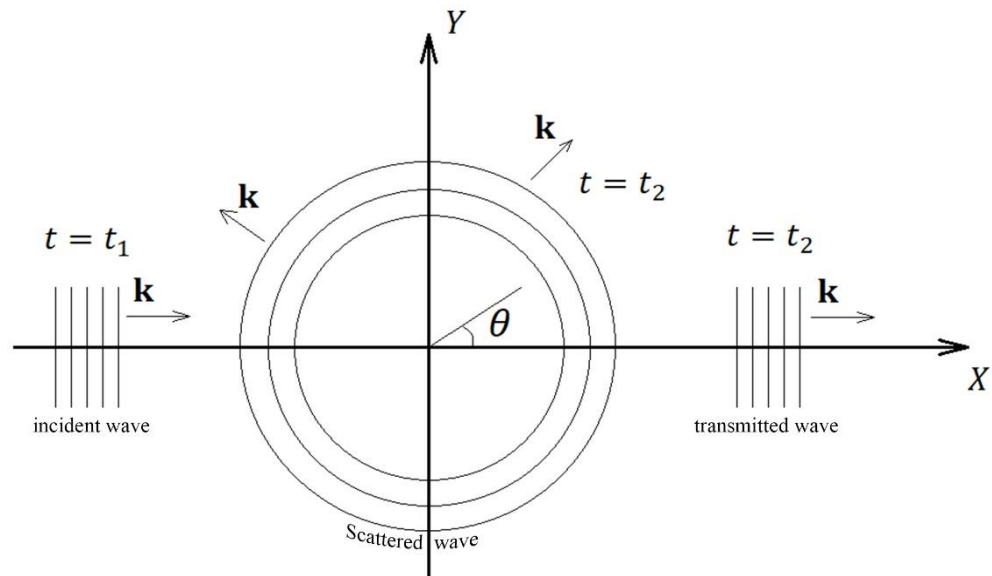

Fig. 2

The Lagrangian of a particle in the presence of a magnetic field is:

$$L_1(\mathbf{x}, \dot{\mathbf{x}}) = L_0(\mathbf{x}, \dot{\mathbf{x}}) + \frac{e}{c}(\mathbf{A}\dot{\mathbf{x}}) \qquad (5)$$

It follows that

$$\int_{t_1}^{t_2} L_1(\mathbf{x}, \dot{\mathbf{x}})\, dt = \int_{t_1}^{t_2} L_0(\mathbf{x}, \dot{\mathbf{x}})\, dt + \int_{x_1}^{x_2} \frac{e}{c}(\mathbf{A}d\mathbf{x}) \qquad (6)$$

Since outside the magnetic filament (an infinite solenoid) the magnetic field **H** is zero, and rot **A** = **H**, then the second term in (5) is the same for the integrals over all paths which go around the magnetic filament from whatever chosen side.

We denote by:

$$\varphi_R = \int_{x_1}^{x_2} \frac{e}{c}(\mathbf{A}d\mathbf{x})$$

the integral taken over the path that bends round the filament on the right, and

$$\varphi_L = \int_{x_1}^{x_2} \frac{e}{c}(\mathbf{A}d\mathbf{x})$$

over the path on the left.

Clearly, if $x_1$ and $x_2$ are pairwise the same for the first and second integrals, then $\varphi = \varphi_R - \varphi_L$ is independent of $x_1$, $x_2$, and

$$\varphi = \varphi_R - \varphi_L = \oint \frac{e}{c}(\mathbf{A}d\mathbf{x}) \qquad (7)$$

where integration is performed counterclockwise over any contour which surrounds the magnetic filament.

By separating in the path integral for the evolution operator (2) those terms which correspond to all paths enveloping the filament on the right, and respectively, on the left, we have:

$$U(x_2, t_2, x_1, t_1) = U_R(x_2, t_2, x_1, t_1)e^{i\varphi_R} + U_L(x_2, t_2, x_1, t_1)e^{i\varphi_L} \tag{8}$$

where, for example,

$$U_R(x_2, t_2, x_1, t_1) = \int_{(x_1,t_1)}^{(x_2,t_2)} \exp\left(i \int_{t_1}^{t_2} L_0(\mathbf{x}, \dot{\mathbf{x}})\, dt\right) D\mathbf{x}(t)$$

is the integral over the paths of a <u>free</u> particle which envelops the filament from the right.

In the absence of a magnetic filament, there is no scattered wave. Thus, if we take the initial wave function in the form (4), then for sufficiently large times $t_2$ we can assume that

$$(U\psi_1)(x_2) = (U_R\psi_1)(x_2) + (U_L\psi_1)(x_2)$$

is close to zero everywhere except for a narrow layer near the plane $(X, Z)$, that is at scattering angles $\theta$ close to zero. Hence everywhere except for the region near the plane $(X, Z)$

$$(U_R\psi_1)(x_2) + (U_L\psi_1)(x_2) = 0 \tag{9}$$

By choosing sufficiently small initial dimensions $d$ and $l$ of the incident wave packet, we may assume that $\varphi_R$ and $\varphi_L$ in (8) are constant for each given scattering angle. Then the amplitude of scattering through the angle $\theta$ will be proportional to

$$f(\theta) \propto (U_R\psi_1)(\theta)e^{i\varphi_R} + (U_L\psi_1)(\theta)e^{i\varphi_L} \tag{10}$$

and the differential scattering cross-section, having regard to (9), will be proportional to

$$|f(\theta)|^2 \propto 2|(U_R\psi_1)(\theta)|^2 - 2|(U_R\psi_1)(\theta)|^2 \cos(\varphi_L - \varphi_R)$$

that is

$$|f(\theta)|^2 \propto \sin^2\frac{\varphi}{2} \tag{11}$$

It can be noted that this result remains valid under assumptions which are more general than those of Aharonov-Bohm – the shape of the filament can be arbitrary, not necessarily straight, sufficing the filament's deviations from the Z axis be limited by some value $r_{max}$. In this case,

(11) remains valid, the scattering cross-sections will still contain the factor $\sin^2 \frac{\varphi}{2}$, although the angular distribution of the scattered particles will differ from that calculated by Aharonov-Bohm, and moreover, in general terms the scattering will lose translational symmetry along the $Z$ axis.

In short, this is the result of the "classical" consideration of the Aharonov-Bohm experiment. It is classical in the sense that the magnetic field created by the solenoid (although everywhere zero except on an infinitely thin filament) is understood as being predetermined. In reality, a magnetic field is not a mathematical abstraction, but is created by moving electric charges. In our case – by charges which move inside the magnetic filament. Moreover, generally this movement has a quantum nature.

The charges moving inside the solenoid move, in particular, under the influence of a magnetic field of a particle moving past the solenoid. We must examine this more closely. If the solenoid is regarded as a quantum system, it will be characterised by its dynamic variables.

We denote their totality by $q$. The Lagrangian of "the solenoid plus particle" system appears thus

$$L(q, \mathbf{x}) = L_s(q, \dot{q}) + L_{int}(q, \dot{q}, \mathbf{A}) + L_0(\mathbf{x}, \dot{\mathbf{x}}) \tag{12}$$

Here $L_s$ is the Lagrangian in the absence of a passing particle's magnetic field, $L_{int}$ is an addition to the Lagrangian arising from the presence of the vector potential $\mathbf{A}$ of a passing particle, and $L_0$ is the Lagrangian of a free particle (3). The transition from considering the motion of a particle in the field of a solenoid to considering the evolution of a solenoid in the field of a particle is legitimate if the retardation of the electromagnetic field is ignored.

For the evolution operator we have the expression

$$U(q_2, x_2, t_2, q_1, x_1, t_1) =$$
$$= \int_{(q_1, x_1, t_1)}^{(q_2, x_2, t_2)} \exp\left(i \int_{t_1}^{t_2} \left(L_s(q, \dot{q}) + L_{int}(q, \dot{q}, \mathbf{A})\right.\right. \tag{13}$$
$$\left.\left. + L_0(\mathbf{x}, \dot{\mathbf{x}})\right) dt\right) Dq(t) D\mathbf{x}(t)$$

We integrate this expression over $Dq(t)$ to obtain:

$$U(q_2, x_2, t_2, q_1, x_1, t_1) = \int_{(x_1,t_1)}^{(x_2,t_2)} U_s(q_2, t_2, q_1, t_1|\mathbf{A}) \exp\left(i \int_{t_1}^{t_2} L_0(\mathbf{x}, \dot{\mathbf{x}}) dt\right) D\mathbf{x}(t) \qquad (14)$$

We denote by $U_s(q_2, t_2, q_1, t_1|\mathbf{A})$ the evolution operator of the solenoid as a functional of the entire function $\mathbf{A}(t)$, $t_1 \leq t \leq t_2$ of the vector potential of the passing particle's magnetic field. (In strict terms, we should use the expression $\mathbf{A}(t, k)$, where $k$ is the index which numbers the entire set of the solenoid's variables, since $\mathbf{A}$ varies at different points of the solenoid. However, for simplicity we omit this index until it is clearly necessary for the purposes of calculation.)

As long as the initial states of the solenoid and the particle <u>are not entangled</u>, that is, the wave function of the initial state of "the solenoid plus particle" system $\psi_1(q, \mathbf{x})$ can be represented as

$$\psi_1(q, \mathbf{x}) = \psi_{s,1}(q)\psi_1(\mathbf{x}) \qquad (15)$$

then it is possible, acting by the evolution operator in the form (14) on the initial state (15), to integrate over $q_1$ to obtain

$$\psi_2(q_2, x_2) = \int dx_1 \int_{(x_1,t_1)}^{(x_2,t_2)} \psi_{s,2}(q_2|\mathbf{A}) \exp\left(i \int_{t_1}^{t_2} L_0(\mathbf{x}, \dot{\mathbf{x}}) dt\right) D\mathbf{x}(t) \cdot \psi_1(x_1) \qquad (16)$$

Here, $\psi_{s,2}(q|\mathbf{A})$- is the result of solving the Schrödinger equation for the entire set of solenoid variables given the solenoid's initial state $\psi_{s,1}(q)$ and the time dependence of the vector potential $\mathbf{A}(t)$ on the interval $[t_1, t_2]$, which corresponds to the given trajectory $x(t)$ of the passing particle.

We prove that at certain assumptions regarding "the internal structure" of the solenoid, the following will be fulfilled:

$$\psi_{s,2}(q|\mathbf{A}) = \psi_{s,2}(q) e^{i\varphi_R}$$

for particle trajectories passing round the solenoid on the right; and

$$\psi_{s,2}(q|\mathbf{A}) = \psi_{s,2}(q) e^{i\varphi_L}$$

for the particle trajectories passing round the solenoid on the left, and $\varphi_L$ and $\varphi_R$ are the same as those which appear in expression (7). In other words: the solenoid's final state which we would obtain in the absence of a passing particle $\psi_{s,2}(q)$, is multiplied, in the event of a particle passing the solenoid, by a phase factor which depends solely on which side the particle passes.

Before starting to prove this statement, we note that if it is true, then we obtain precisely the Aharonov-Bohm result. Indeed, in this event the path integral in (16) is divided into two terms, corresponding to the passing of the solenoid on the right and left sides, in each of which $\psi_{s,2}(q)$ is removed from the integral, and we thus have

$$\psi_2(q_2, x_2) = \psi_{s,2}(q)\left(e^{i\varphi_R}(U_R\psi_1)(x_2) + e^{i\varphi_L}(U_L\psi_1)(x_2)\right) \tag{17}$$

This expression, up to the factor $\psi_{s,2}(q)$, of the wave function of the solenoid's final state in the absence of a passing particle, corresponds to expression (8) for the operator of particle evolution in the solenoid field. However, the factor $\psi_{s,2}(q)$ is plainly irrelevant for calculating a particle's scattering cross-section.

We consider a (not necessarily straight) magnetic filament formed by a large number of identical magnetic dipoles. Let these dipoles be particles with spin $1/2$ and magnetic moment **μ**, located in a filament with the same density $n$, and oriented along the tangent to the filament at each of its points.

We define Pauli matrices $\sigma_x$, $\sigma_y$, $\sigma_z$ and states $\xi = \begin{pmatrix}1\\0\end{pmatrix} = |\uparrow\rangle$ and $\eta = \begin{pmatrix}0\\1\end{pmatrix} = |\downarrow\rangle$ for each spin such that with the Z axis directed along the (oriented) tangent to the filament $\sigma_z$ is diagonal.

Let the initial state of the filament at time $t_1$ in the specified basis be:

$$|\psi_{s,1}\rangle = |\uparrow_1\rangle|\uparrow_2\rangle \ldots |\uparrow_k\rangle \ldots |\uparrow_N\rangle \tag{18}$$

where the index $k$ numbers all the dipoles in the filament.

We will assume that the unperturbed Hamiltonian of the filament equals the sum of the unperturbed Hamiltonians of the dipoles, which for each dipole have the form:

$$H_0 = -\epsilon\sigma_z \tag{19}$$

The corresponding $\sigma_z$ is defined in the state space of each dipole, and this definition, in general terms, depends on the dipole's location on the filament. The fact that this Hamiltonian will have not only positive eigenvalues is in fact a consequence of the displacement of the energy origin, and does not affect the final results in relation to our calculations. The choice of the dipole Hamiltonian in the form of (19) and of the initial state in the form of (18) physically corresponds to the mean field approximation for a ferromagnet, which we assume to be the source of the solenoid magnetization.

The charge which passes the filament creates a magnetic field **h** at the point where the dipole is located. In general terms, its direction does not coincide with the dipole's orientation. The unperturbed Hamiltonian (19) is substituted by

$$H = -\epsilon \sigma_z - \mu(\mathbf{h}\boldsymbol{\sigma}) \tag{20}$$

We will assume that

$$\epsilon \gg \mu h \tag{21}$$

that is, the magnetic field of the passing particle is a small perturbation for the dipole, and

$$\epsilon \gg 1/\tau \tag{22}$$

where $\tau$ is the typical time for a particle to go past the filament.

The condition in (22) means that the adiabatic approximation is applicable. The correspondence of the conditions in (21) and (22) to real experiments will be considered below.

In the first order of perturbation theory, the eigenfunctions of Hamiltonian (20) coincide with $\xi = \begin{pmatrix}1\\0\end{pmatrix}$ and $\eta = \begin{pmatrix}0\\1\end{pmatrix}$, and the eigenvalues become equal to:

$$\epsilon_1 = -\epsilon - \mu h_z, \qquad \epsilon_2 = \epsilon + \mu h_z \tag{23}$$

where $h_z$ is the projection of the magnetic field onto the tangent to the filament at the point where the dipole is located.

In the adiabatic approximation, the solution to the Schrödinger equation with Hamiltonian (20) and the state at the initial moment $t_1$

$$\psi(t_1) = c_1 \xi + c_2 \eta \tag{24}$$

will be:

$$\psi(t) = c_1 \exp\left(-i \int_{t_1}^{t} \epsilon_1(t) dt\right) \xi + c_2 \exp\left(-i \int_{t_1}^{t} \epsilon_2(t) dt\right) \eta \tag{25}$$

Substituting in (25) $\epsilon_1$ and $\epsilon_2$ from (23) we have:

$$\psi(t) = c_1 e^{i\epsilon(t-t_1)} \exp\left(i \int_{t_1}^{t} \mu h_z(t) dt\right) \xi + c_2 e^{-i\epsilon(t-t_1)} \exp\left(-i \int_{t_1}^{t} \mu h_z(t) dt\right) \eta \tag{26}$$

Taking into account (18) we are at this point concerned with the case where $c_1 = 1$, $c_2 = 0$, and then

$$\psi(t_2) = e^{i\epsilon(t_2-t_1)} \exp\left(i \int_{t_1}^{t_2} \mu h_z(t) dt\right) \xi \tag{27}$$

Using (18) as the initial state of the filament, we obtain for the state at time $t_2$

$$|\psi_s(t_2|A)\rangle = |\psi_s(t_2)\rangle \exp\left(i \sum_k \int_{t_1}^{t_2} \mu h_{z,k}(t) dt\right) \tag{28}$$

where the notation $|\psi_s(t_2|A)\rangle$ is used to emphasise that the evolution of the filament in a magnetic field is being considered, $|\psi_s(t_2)\rangle$ is the state into which the filament would pass in the absence of the magnetic field, and $h_{z,k}$ is the projection of the magnetic field onto the tangent to the filament at the location of the dipole with the number $k$.

We transform the exponent in (28) within the limit of a large number of dipoles:

$$\sum_k \mu h_{z,k} = \int_S dl \cdot n \cdot \mu \cdot h_z(l) \tag{29}$$

where $n$ is the linear density of the number of dipoles, and l- is the natural parameter on the $S$ curve describing the filament. Integration is performed over this entire curve. In turn,

$$\int_S dl \cdot n \cdot \mu \cdot h_z(l) = \int_S \mu n(\mathbf{hdl}) \tag{30}$$

Recalling that the magnetic field of a moving particle at the location of a dipole is

$$\mathbf{h} = \frac{e}{c} \frac{[\dot{\mathbf{x}} \times \mathbf{R'}]}{R'^3} \tag{31}$$

where $\dot{\mathbf{x}}$ is the velocity of the particle and $\mathbf{R'}$ is the radius vector from the particle to the location of the dipole, we have:

$$\int_{t_1}^{t_2} dt \int_S \mu n (\mathbf{h} d\mathbf{l}) = \int_{t_1}^{t_2} dt \int_S \mu n \frac{e}{c} \left( \frac{[\dot{\mathbf{x}} \times \mathbf{R'}]}{R'^3} d\mathbf{l} \right) \tag{32}$$

We substitute here $\mathbf{R'}$ by $-\mathbf{R}$, where $\mathbf{R}$ is the radius vector from <u>the dipole location to the particle</u>, and rearrange the factors in the mixed product.

We then have:

$$\int_{t_1}^{t_2} dt \int_S \mu n \frac{e}{c} \left( \frac{[\dot{\mathbf{x}} \times \mathbf{R'}]}{R'^3} d\mathbf{l} \right) = \int_S \int_{t_1}^{t_2} \frac{e}{c} \left( \frac{[\mu n d\mathbf{l} \times \mathbf{R}]}{R^3} \dot{\mathbf{x}} \right) dt$$

But the integral over the filament equals to the vector potential of its magnetic field at the point $\mathbf{R}$ of the particle's location:

$$\int_S \frac{[\mu n d\mathbf{l} \times \mathbf{R}]}{R^3} = \int_S \frac{[d\mathbf{m} \times \mathbf{R}]}{R^3} = \mathbf{A}(\mathbf{R}) \tag{33}$$

Here, $d\mathbf{m}$ denotes the magnetic moment of the filament's section $d\mathbf{l}$. Plainly, $d\mathbf{m} = \mu n d\mathbf{l}$, and the final equality in (33) simply expresses the fact that the vector potential created by the filament equals the sum of the vector potentials in the dipoles of which it is composed. Taking account of (33) we then have, for the exponent in (28), the following expression:

$$\sum_k \int_{t_1}^{t_2} \mu h_{z,k}(t) dt = \int_{x_1}^{x_2} \frac{e}{c} (\mathbf{A}(\mathbf{x}) d\mathbf{x}) \tag{34}$$

and (28) can be rewritten as:

$$|\psi_s(t_2|A)\rangle = |\psi_s(t_2)\rangle \exp\left( i \frac{e}{c} \int_{x_1}^{x_2} (\mathbf{A} d\mathbf{x}) \right) \tag{35}$$

where $x_1, x_2$ are the initial and final positions of the passing particle. Here, on the left-hand side "$A$" <u>symbolises</u> the vector potential of the passing particle, in which the solenoid is located, and on the right-hand side **A** denotes the vector potential of the solenoid on the trajectory of this particle. Formula (35) is precisely what we set out to prove.

Naturally, it would be remarkably naïve to suppose that somewhere, even in the most advanced nations, a solenoid with the initial state (18) could ever be produced. However, we are able to prove a statement similar to that of (17) and under weaker assumptions about the structure of the solenoid. That is to say, the initial state of the elementary dipoles of the solenoid can look like the superposition

$$\psi = c_1 \xi + c_2 \eta \tag{36}$$

and, moreover, be entangled with the environment:

$$\psi = c_1 \xi \otimes \chi + c_2 \eta \otimes \nu \tag{37}$$

where $\chi$, $\nu$ are vectors in the state space of the environment. That is, in principle, solenoid dipoles can be described by an arbitrary density matrix. It is understood that the Hamiltonian remains in the form (20), and the conditions (21) and (22) remain valid. Since we are now uncertain as to the accuracy of (17), to calculate the scattering cross-section we square the modulus of the wave function $\psi_2(q_2, x_2)$ as given in the formula (16), and sum over the variables $q$:

$$|f(\theta)|^2 \propto \int |\psi(q_2, x_2)|^2 dq_2 =$$

$$= \int dq_2 \left| \int dx_1 \int_{(x_1,t_1)}^{(x_2,t_2)} \psi_{s,2}(q_2|\mathbf{A}) \exp\left(i \int_{t_1}^{t_2} L_0(\mathbf{x}, \dot{\mathbf{x}}) dt \right) D\mathbf{x}(t) \psi_1(x_1) \right|^2 \tag{38}$$

We convert this to the form:

$$\int |\psi(q_2, x_2)|^2 dq_2 =$$

$$= \int dq_2 \int dx_1 \int dy_1 \int_{(x_1,t_1)}^{(x_2,t_2)} \int_{(y_1,t_1)}^{(y_2,t_2)} \psi^*(x_1) \psi(y_1) \psi^*_{s,2}(q_2|\mathbf{A}[x]) \psi_{s,2}(q_2|\mathbf{A}[y]) \times$$

$$\times \exp\left(-i \int_{t_1}^{t_2} L_0(x, \dot{x}) dt \right) \exp\left(i \int_{t_1}^{t_2} L_0(y, \dot{y}) dt \right) Dx(t) Dy(t) \tag{39}$$

Here, $\psi_{s,2}(q_2|\mathbf{A}[x])$ and $\psi_{s,2}(q_2|\mathbf{A}[y])$ denote the wave functions of the states of the solenoid, into which it transfers, being in the time interval $[t_1, t_2]$ in the field of a charge moving along the trajectories $x(t)$ и $y(t)$, respectively.

Changing the order of integration, we find in (39) the factor

$$\langle \psi^*_{s,2}(\mathbf{A}[x]) | \psi_{s,2}(\mathbf{A}[y]) \rangle = \int dq_2 \, \psi^*(q_2|\mathbf{A}[x])\psi(q_2|\mathbf{A}[y]) \tag{40}$$

Since the initial state of the solenoid is now a direct product of the states of individual dipoles of the form (36) or (37), and the Hamiltonian (20) does not entangle them, then $|\psi_{s,2}(\mathbf{A}[x])\rangle$ will likewise be the direct product of the states into which states (36) or (37) of individual dipoles evolve with Hamiltonian (20). An explicit expression for these states is the formula (26). Where the initial state is entangled with the environment, it is necessary to substitute $\xi$ and $\eta$ in (26) by $\xi \otimes \chi$ and $\eta \otimes \nu$ from (36). Since, as will soon become clear, only the <u>orthogonality</u> of vectors in the two terms of (26) is important for our calculations, we will no longer consider the differences between (36) and (37). We divide the solenoid into sections so small that the field $h_z$ in them can be considered the same for all the dipoles they contain, yet at the same time large enough that there are macroscopically many such dipoles in a section.

Evidently,

$$|\psi_{s,2}(\mathbf{A}[x])\rangle = \prod_{l=1}^{L} |\psi_l(\mathbf{A}[x])\rangle \tag{41}$$

where $L$ is the number of such sections.

In turn, the wave function of the $l$-th section:

$$|\psi_l(\mathbf{A}[x])\rangle = \prod_{k=1}^{K_l} (c_1(t)\xi_k + c_2(t)\eta_k) \tag{42}$$

where $K_l$ is the number of dipoles in the l-th section, and $c_1(t)$ and $c_2(t)$ are the coefficients at the state vectors in (26):

$$c_1(t) = c_1 e^{i\epsilon(t-t_1)} \exp\left(i \int_{t_1}^{t} \mu h_z(t) dt\right), \quad c_2(t) = c_2 e^{-i\epsilon(t-t_1)} \exp\left(-i \int_{t_1}^{t} \mu h_z(t) dt\right)$$

Further, it is apparent that

$$|c_1(t)|^2 = |c_1|^2, \quad |c_2(t)|^2 = |c_2|^2, \quad |c_1|^2 + |c_2|^2 = 1 \tag{43}$$

We consider an arbitrary set of $n$ spins $1/2$ with base states $\xi, \eta$.

Consider, then, the state of this set, in which certain m of the spins are in the $\eta$ state ("inverted"), and the rest of the $n - m$ spins are in the $\xi$ state. There will be a total of $C_n^m$ such states. We denote the sum of all these states as $\left|\binom{\xi^n}{\eta^m}\right\rangle$

(the designation of the binomial coefficient $C_n^m$ in the form $\binom{n}{m}$ is taken as a model)

$\left|\binom{\xi^n}{\eta^m}\right\rangle$ has two easily-proved properties:

$$\prod_{k=1}^{n}(a\xi_k + b\eta_k) = \sum_{m=0}^{n} a^m b^{n-m} \left|\binom{\xi^n}{\eta^m}\right\rangle \tag{44}$$

and

$$\left\|\binom{\xi^n}{\eta^m}\right\|^2 = C_n^m \tag{45}$$

Taking into account (44) and (45)

$$\langle \prod_{k=1}^{n}(a\xi_k + b\eta_k) | \prod_{k=1}^{n}(a_1\xi_k + b_1\eta_k)\rangle = \sum_{m=0}^{n}(a^*a_1)^m (b^*b_1)^{n-m} C_n^m \tag{46}$$

Using (46) to calculate $\langle\psi_l(\mathbf{A}[x])|\psi_l(\mathbf{A}[y])\rangle$, we obtain:

$$\langle\psi_l(\mathbf{A}[x])|\psi_l(\mathbf{A}[y])\rangle =$$
$$= \sum_{m=0}^{K_l} C_{K_l}^m (|c_1|^2)^m (|c_2|^2)^{K_l-m} \times \tag{47}$$
$$\times \exp(-im\Delta\varphi[x] + im\Delta\varphi[y] + i(n-m)\Delta\varphi[x] - i(n-m)\Delta\varphi[y])$$

where $\Delta\varphi[x]$ denotes the phase addition arising due to the adiabatic shift of dipole energy levels in the magnetic field of a particle moving along the trajectory $x(t)$:

$$\Delta\varphi[x] = \int_{t_1}^{t_2} \mu h_z(t) dt$$

Since, to the extent that the particle trajectory is distant from the dipole,

$$\Delta\varphi[x], \Delta\varphi[y] \ll 2\pi \tag{48}$$

then the exponential in (47) varies very slowly with a change in $m$ in comparison with the pre-exponential term. Further, since $K_l \gg 1$ and $|c_1|^2 + |c_2|^2 = 1$, then $m$ and $(n-m)$ under the exponent in (47) can be substituted by their "mean values" $|c_1|^2 K_l$ and $|c_2|^2 K_l$. Hence we have:

$$\langle\psi_l(\mathbf{A}[x])|\psi_l(\mathbf{A}[y])\rangle = \exp\left(iK_l((-|c_1|^2 + |c_2|^2)\Delta\varphi[x] + (|c_1|^2 - |c_2|^2)\Delta\varphi[y])\right) \tag{49}$$

But

$$K_l(|c_1|^2 - |c_2|^2)\mu = \overline{m_l} \tag{50}$$

where $\overline{m_l}$ is the magnetic moment of the $l$-th section of the solenoid. Thus,

$$\langle\psi_l(\mathbf{A}[x])|\psi_l(\mathbf{A}[y])\rangle = \exp(-i\Delta\varphi_l[x] + i\Delta\varphi_l[y]) \tag{51}$$

where

$$\Delta\varphi_l[x] = \int_{t_1}^{t_2} \overline{m_l}\, h_{z,l} dt = \int_{t_1}^{t_2} (\overline{\mathbf{m}_l}\mathbf{h}_l)\, dt \tag{52}$$

where $\mathbf{h}_l, h_{z,l}$ denote the vector of the magnetic field of a particle in the region of the $l$-th section of the solenoid, and, respectively, its longitudinal component.

It follows that

$$\langle\psi_{s,2}(\mathbf{A}[x])|\psi_{s,2}(\mathbf{A}[y])\rangle = \prod_l \langle\psi_l(\mathbf{A}[x])|\psi_l(\mathbf{A}[y])\rangle = \\ = \exp\left(-i\sum_{l=1}^{L}\Delta\varphi_l[x] + i\sum_{l=1}^{L}\Delta\varphi_l[y]\right) \tag{53}$$

Passing here from summation over $l$ to integration over the solenoid, we have

$$\langle\psi_{s,2}(\mathbf{A}[x])|\psi_{s,2}(\mathbf{A}[y])\rangle \\ = \exp\left(-i\int_{t_1}^{t_2}\int_S n\bar{\mu}\,(\mathbf{h}([x],t))\mathrm{dl})\mathrm{d}t + i\int_{t_1}^{t_2}\int_S n\bar{\mu}\,(\mathbf{h}([y],t))\mathrm{dl})\mathrm{d}t\right) \tag{54}$$

where $\bar{\mu} = (|c_1|^2 - |c_2|^2)\mu$ is the quantum average of the projection of the dipole's magnetic moment on the $Z$ axis, and $\mathbf{h}([x], t)$ is the magnetic field of a particle passing along the trajectory $x(t)$.

Comparing this with (28), (29), and (30), we conclude that where the solenoid dipoles are in an arbitrary superposition of states, as well as in a mixed state, the wave functions of the solenoid state under the action of the magnetic field of a passing particle will behave in the same way as in the case where the dipoles are oriented strictly along the solenoid axis. That is to say, the initial state of the solenoid $|\psi_{s,1}\rangle$ after a particle has moved past it along the trajectory $x(t)$ will move into the state $|\psi_{s,2}(\mathbf{A}[x])\rangle$, which <u>for the purposes of calculating the scalar product with another similar state</u> $|\psi_{s,2}(\mathbf{A}[y])\rangle$, may simply be assumed as equal to

$$|\psi_{s,2}(\mathbf{A}[x])\rangle = \exp\left(i \int_{t_1}^{t_2} dt \int_S n\bar{\mu}(\mathbf{h}d\mathbf{l})\right) |\psi_{s,1}\rangle \tag{55}$$

where $\bar{\mu}$ is the quantum-mechanical average of the projection of the dipole's magnetic moment onto the solenoid axis. We emphasise that this fact is essentially a consequence of the quantum-statistical properties of a large number of dipoles.

From this point onwards, we can repeat verbatim all the calculations given in formulas (31) - (35), except that in them $\mu$ has to be substituted by $\bar{\mu}$.

Having proved the effectiveness of (35), integration over $Dx(t)$ and $Dy(t)$ in (39) can each be divided into two regions: over paths which pass the filament on the left, and respectively, on the right. In virtue of (35), in each such region the factors $|\psi_{s,2}\rangle$ and $\langle\psi_{s,2}|$ can be removed from the integral over $Dx(t)$ and $Dy(t)$, and evidently, when calculating the scattering cross-section we again obtain the factor $\sin^2\frac{\varphi}{2}$.

This may be considered as complete proof that the Aharonov-Bohm effect can be derived not as a result of the vector potential of the solenoid on a particle in those regions where the magnetic field created by the solenoid is zero, but as a result of the effect of a completely real nonzero magnetic field of the particle on the quantum-mechanical objects that make up the solenoid.

## Experimental verification of the action of the shielded potential

Does the above reasoning have any value apart from the fact that it will provide satisfaction to some and displeasure to others through the complete and final elimination of the principles of nonlocality from the explanation of the Aharonov-Bohm effect? Isn't this reasoning simply a different method of reasoning, which does not lead to any further consequences? Of course not. Our conclusion begins with the fact that the solenoid which causes the Aharonov-Bohm effect is material. In the case of actual experiments, this materiality is manifested, for example, in the fact that this solenoid must have very small (~1 μm) dimensions, and interact with a particle which passes it at similarly small distances, in respect of which the question of how a passing particle interacts with each quantum component of the solenoid becomes fundamental. In particular, proceeding from the foregoing, we can state with certainty that the application of perturbation theory and adiabaticity are not assumptions which simplify our proof, but fundamental conditions for the Aharonov-Bohm effect to manifest itself. Experiments which violate these conditions will not produce a picture which corresponds to what is usually expected from the Aharonov-Bohm effect.

In addition, as an example of the effectiveness of the approach submitted here, consider the recent work of Saldanha [9], which treats the indestructibility of the quantum phase. As an argument to confirm their conclusions, the authors cite the results of an experiment by Tonomura et al. [10, 11], in which the Aharonov-Bohm effect was successfully demonstrated for a solenoid surrounded by a superconducting shield. From the standpoint of Saldanha [9], the Aharonov-Bohm effect should appear, regardless of whether the solenoid is surrounded by a superconducting shield.

Our method of consideration easily leads to the conclusion that if the solenoid is surrounded by a shield that completely isolates it from the magnetic field of a passing particle, the Aharonov-Bohm effect will not be observed, since the contribution of all the perturbations we have considered towards evolution of the quantum constituent parts of the solenoid is plainly zero. It follows that something is amiss with the reasoning in [9], but that is not the subject of our consideration. The following questions are much more interesting:

1. What does an experiment with a shielded solenoid look like in terms of the "classical" theory of the Aharonov-Bohm effect?

2. Why, in spite of our statements about the absence in this case of the Aharonov-Bohm effect, did the Tonomura experiment give such an excellent result?

The answer to the first question is that the currents arising in the shield under the action of the field of a passing particle and shielding that field are, at the same time, in the vector potential of the solenoid, from which they acquire a quantum phase, which completely compensates for the phase acquired in the vector potential of the solenoid by a passing charge.

Proof of this is given in the Appendix, in which an appraisal is made with certain restrictions on the form of the shield, although it is plain that there should be a proof for the general case.

For the second question, as to why the Tonomura experiment achieved success, there is a simple answer – the superconducting shield in his case did not fulfil its function, that is, it did not shield the field of the passing particle.

As described by Tonomura [10, 11], the energy of passing electrons in his experiments was 150 keV, which corresponds to a velocity of $\sim 0.77$ c. In turn, the time during which an electron passed the solenoid, that is, travelled a distance of $\sim 1$ μm, was $\sim 10^{-14}$ sec. It is known from [22–24] that no superconductor is capable of shielding radiation with a frequency of $\sim 10^{14}$ Hz, which corresponds to the infrared region of the spectrum. In reality, the shielding region of superconductors is limited by frequencies:

$$\hbar\omega \lesssim 3.5 k T_c$$

where $T_c$ is the critical temperature of the superconductor [22]. In the case under consideration (superconducting niobium), this corresponds to frequencies of the order of hundreds of gigahertz (600 GHz). Above these frequencies, a niobium film 0.25 μm thick, as in Tonomura's experiment, would not be able to shield anything, let alone the field of a passing electron.

As an incidental result of our reasoning, we have a prediction that an experiment to detect the Aharonov-Bohm effect using a shielded solenoid, which would use sufficiently slow electrons (with an energy of the order of several eV), would produce a negative result.

## Appendix

## The action of a shielded solenoid on passing particles

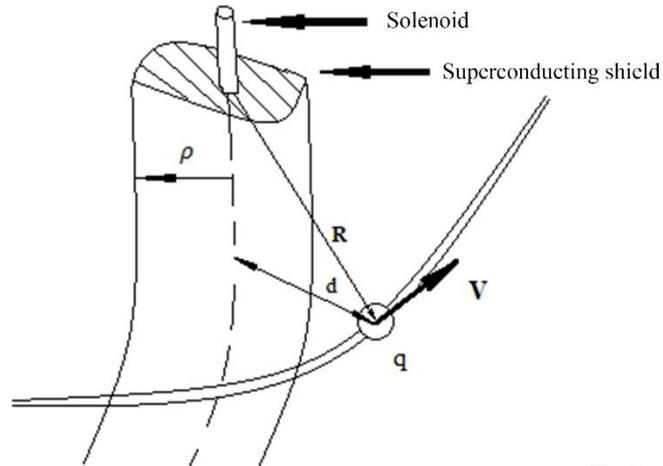

Fig. 3

The solenoid is assumed to be infinitely thin, and the superconducting shield to be very close to the solenoid as shown in Fig. 3, so that the distance $d$ from the charge to the nearest point of the solenoid is much greater than the size $\rho$ of the shield cross-section by a plane perpendicular to the tangent to the solenoid: $d \gg \rho$, so that the magnetic field of a passing charge can be considered constant throughout the given cross-section.

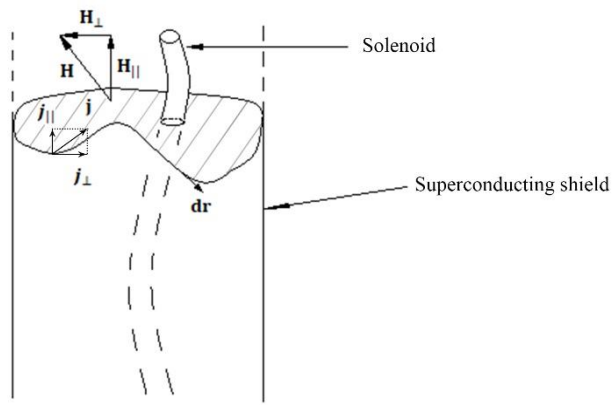

Fig. 4

In addition, we assume that $\rho \ll K$ (the solenoid's radius of curvature). We divide the magnetic field **H**, created by a passing charge in a certain section, into $\mathbf{H}_\parallel$ – the component parallel to the tangent to the solenoid, and $\mathbf{H}_\perp$ – the component parallel to the section plane as shown in Fig. 4. If the currents flowing in the shield completely compensate for the magnetic field created by the passing charge, then it is evident that

$$H_{||} = -\frac{4\pi}{c} j_\perp \qquad (A.1)$$

where $j_\perp$ is the component of the (surface) current density which lies in the cross-sectional plane. The value $j_\perp$ is constant along the entire perimeter of the solenoid cross-section.

Because of the interaction of the currents induced in the superconducting shield with the vector potential of the solenoid's magnetic field, in a short time $\delta t$ the elementary charge carriers which create these currents (in this case, Cooper pairs) acquire an aggregate additional phase of their common wave function:

$$\delta\varphi = \int (\mathbf{j} \cdot \mathbf{A})\, ds\, \delta t \qquad (A.2)$$

where the integral is taken over the entire surface of the superconducting shield.

Taking into account that the shield is close to the solenoid, vector **A** lies in the plane of the shield's cross-section, and hence:

$$\delta\varphi = \int (\mathbf{j}_\perp \cdot \mathbf{A})\, ds\, \delta t \qquad (A.3)$$

and this, in turn, can be converted into the expression:

$$\delta\varphi = \int dl\, j_\perp \oint (\mathbf{dr} \cdot \mathbf{A})\, \delta t \qquad (A.4)$$

where the integral over $dl$ is taken over the solenoid's entire length, and the integral over **dr** is taken over the cross-section perimeter at the current point.

But

$$\oint (\mathbf{dr} \cdot \mathbf{A}) = \Phi \qquad (A.5)$$

where $\Phi$ is the magnetic flux passing through the circuit of integration. In this case, $\Phi$ is constant along the entire length of the shield, and is equal to the magnetic flux passing inside the solenoid.

Hence,

$$\delta\varphi = \Phi \int j_\perp dl \delta t \tag{A.6}$$

It follows from (1) that $j_\perp \, dl = -\frac{c}{4\pi} H_\parallel dl = -\frac{c}{4\pi} \frac{q}{c} \frac{([\mathbf{V} \times (-\mathbf{R})] \cdot \mathbf{dl})}{R^3}$ that is,

$$j_\perp \, dl = -\frac{q}{4\pi} \frac{([\mathbf{dl} \times \mathbf{R}] \cdot \mathbf{V})}{R^3} \tag{A.7}$$

where $\mathbf{V}$ is the velocity of the charge, and $\mathbf{R}$ – is the vector passing from the current point of the solenoid to the charge, whence it follows that

$$\varphi = \Phi \int j_\perp dl \delta t = -\left( \int \frac{[\mathbf{dl} \times \mathbf{R}]}{R^3} \frac{\Phi}{4\pi} \cdot q\mathbf{V} \right) \delta t \tag{A.8}$$

But

$$\int \frac{[\mathbf{dl} \times \mathbf{R}]}{R^3} \frac{\Phi}{4\pi} = \mathbf{A}(\mathbf{x}) \tag{A.9}$$

where $\mathbf{A}(\mathbf{x})$ is the vector potential created by the solenoid at the point where the charge $\mathbf{x}$ is located.

Hence,

$$\delta\varphi = -q(\mathbf{A} \cdot \mathbf{V})\delta t = -q(\mathbf{A} \cdot \delta\mathbf{x}) \tag{A.10}$$

But $q(\mathbf{A} \cdot \delta\mathbf{x})$ is equal to the phase acquired during the time $\delta t$ by the passing charge. Hence the phases acquired by the superconducting shield and the passing charge mutually cancel each other out at each moment of time.